\documentclass[a4paper,pra,aps,amsmath,amssymb]{revtex4}

\usepackage[T1]{fontenc}
\usepackage[utf8]{inputenc} 
\usepackage[english]{babel}    
\usepackage[pdftex]{graphicx} 
\usepackage{hyperref}         
\usepackage{amsmath}
\usepackage{physics}
\usepackage{xcolor}

\usepackage[columnsep=14.173228pt,textwidth=518.740158pt,left=38.267717pt,top=38.267717pt,bottom=38.267717pt,includefoot]{geometry}

\newcommand{\heff}{\hbar_\text{eff}}

\usepackage{sfmath}
\usepackage{caption}
\DeclareCaptionLabelSeparator{bar}{~|~}
\usepackage{lmodern}
\fontsize{6pt}{11pt}\selectfont

\begin{document}

\onecolumngrid
\begin{center}
\LARGE \textbf{Chaos-assisted tunneling resonances in a synthetic Floquet superlattice}
\end{center}

\begin{center}
M. Arnal$^1$, G. Chatelain$^1$, M. Martinez$^2$, N. Dupont$^1$, O. Giraud$^3$, D. Ullmo$^3$, B. Georgeot$^2$, G. Lemari\'e$^2$, J. Billy$^1$ and D. Gu\'ery-Odelin$^{1*}$
\end{center}
{\small
$^1$ Laboratoire Collisions Agr\'egats R\'eactivit\'e,  IRSAMC, Universit\'e de Toulouse, CNRS, UPS, France \\
$^2$ Laboratoire de Physique Th\'eorique, IRSAMC, Universit\'e de Toulouse, CNRS, UPS, France \\
$^3$ Universit\'e Paris-Saclay, CNRS, LPTMS, 91405, Orsay, France \\ }

\twocolumngrid

\textbf{The field of quantum simulation, which aims at using a tunable quantum system to simulate another, has been developing fast in the past years as an alternative to the all-purpose quantum computer \cite{kendon2010quantum, cirac2012goals, hauke2012can, schaetz2013focus, RevModPhys.86.153, gross2017quantum, bloch2018quantum, PhysRevLett.110.033902}. In particular, the use of temporal driving has attracted a huge interest recently as it was shown that certain fast drivings can create new topological effects \cite{lindner2011floquet, Topology_cold_atoms4, rechtsman2013photonic, Dalibard2019RMP, DalibardPRX, EckardtRMP}, while a strong driving leads to \emph{e.g.} Anderson localization physics \cite{Casati:LocDynFirst:LNP79, Fishman:LocDynAnders:PRL82, PhysRevLett.67.255, Moore:AtomOpticsRealizationQKR:PRL95, Casati:IncommFreqsQKR:PRL89, Chabe:Anderson:PRL08}. In this work, we focus on the intermediate regime to observe a quantum chaos transport mechanism called chaos-assisted tunneling \cite{Tomsovic94, raizen, steck2002fluctuations, dembowski2000first, hofferbert2005experimental, dietz2014spectral, PhysRevLett.104.163902} which provides new possibilities of control for quantum simulation. Indeed, this regime generates a rich classical phase space where stable trajectories form islands surrounded by a large sea of unstable chaotic orbits. This mimics an effective superlattice for the quantum states localized in the regular islands, with new controllable tunneling properties. Besides the standard textbook tunneling through a potential barrier, chaos-assisted tunneling corresponds to a much richer tunneling process where the coupling between quantum states located in neighboring regular islands is mediated by other states spread over the chaotic sea. This process induces sharp resonances where the tunneling rate varies by orders of magnitude over a short range of parameters. We experimentally demonstrate and characterize these resonances for the first time in a quantum system. This opens the way to new kinds of quantum simulations with long-range transport and new types of control of quantum systems through complexity.}

\begin{figure}[h!]
\centering
\includegraphics[width=\linewidth]{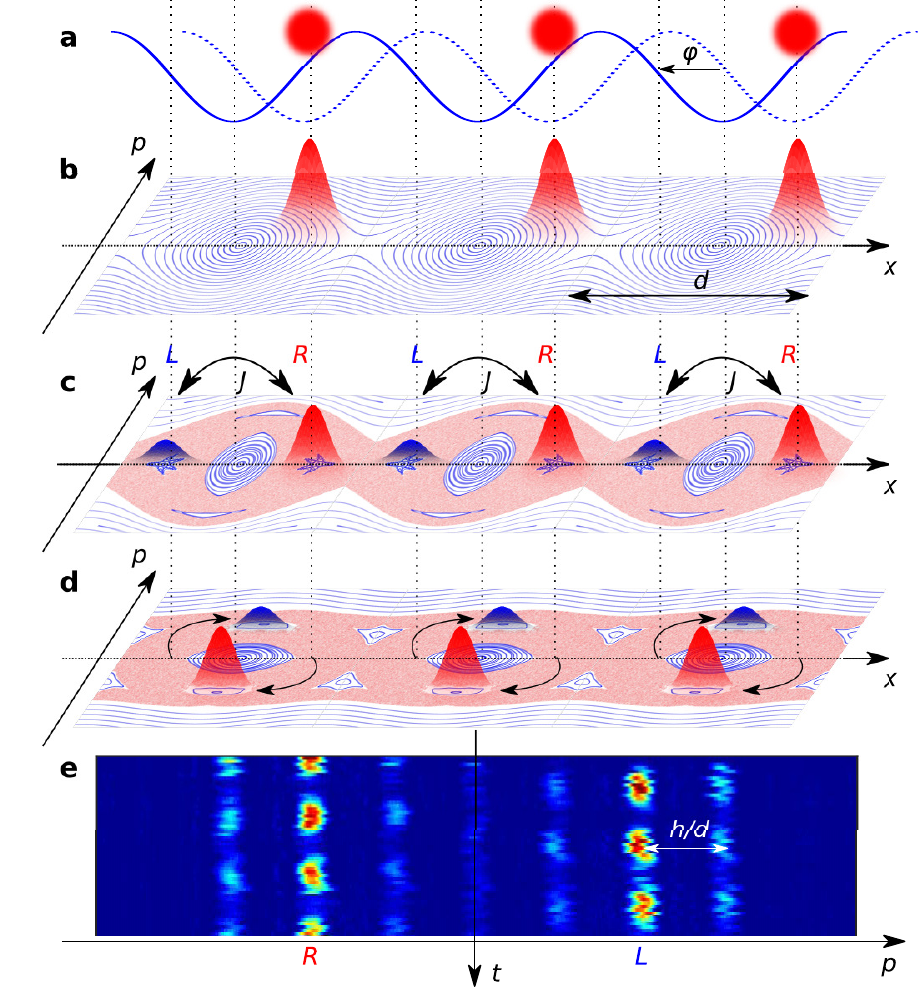}
  \caption{\textbf{Sketch of the experimental procedure}. \textbf{a}, By an abrupt phase shift $\varphi$ of the lattice, the atomic clouds, initially at the bottom of the static lattice wells (dotted blue line), are suddenly placed on the slope of the shifted lattice wells (solid blue line). In the phase space ($x$,$p$) representation shown in (\textbf{b}), it corresponds to a displacement along the $x$-axis. \textbf{c}, The amplitude modulation of the lattice depth generates a mixed phase space with regular islands (blue closed orbits) surrounded by a chaotic sea (red zone). With the appropriate phase shift prior to the modulation, the wave packet can be placed on a lateral regular island (e.g. the right one). Then the wave packet tunnels back and forth between the two stable symmetric lateral islands (tunneling rate $J$), leading to two wave packets $R$ and $L$. \textbf{d}, The observation of the tunneling requires a phase space rotation (black arrows) which transfers to momentum space the information encoded in position space:  the population on the right (resp. left) island gains a negative (resp. positive) momentum. \textbf{e}, Stack of integrated experimental images taken every two modulation periods. The images reflect the momentum distribution after a 25 ms time-of-flight, its periodicity $h/d$ is directly related to the lattice periodicity. Classically, atoms are expected to stay on the same side (initially in the right island). The first three experimental tunneling oscillations are shown.}
  \label{figure1}
\end{figure}

\textit{Introduction.--} 
In classical dynamical systems, a major progress of the last century has been the understanding that many systems display various degrees of chaos \cite{Ott:ChaosDynSyst:02}. Strongly chaotic systems are characterized by exponential sensitivity to initial conditions and ergodicity of a typical trajectory in phase space. Such properties can be seen in very simple systems: in astronomy \cite{laskar1990chaotic}, in chemistry \cite{petrov1993controlling}, in biology \cite{goldberger1990chaos, poon1997decrease} and in ecology \cite{gilpin1979spiral, rinaldi1993multiple}. Most systems exhibit a mixed phase space in which chaotic and regular zones coexist. While chaotic behavior is hard to predict, the instability that emerges in such systems makes them very versatile as a small perturbation can induce a completely different evolution. Even more, this instability can be harnessed to tune the system to a desired final state at a minimal cost, a process dubbed ``control of chaos'' \cite{PhysRevLett.64.1196, boccaletti2000control, ChaosControl}.

In the field of quantum simulation, the emphasis has been put on the use of regular systems which are \emph{a priori} easier to control. Here we show that using chaotic dynamics in a cold atom setting actually opens possibilities for quantum simulations which are difficult to reach by other means. To do so we use periodically driven systems which are a paradigm for the study of quantum chaos as the strength of the modulation can be used to control the level of chaos \cite{chirikov1979universal, Ott:ChaosDynSyst:02, casati2006quantum}. More precisely, we consider ultracold atoms in a 1D optical lattice, whose depth is periodically modulated in time \cite{Mouchet01, DubertrandCAT}.  
For a moderately strong modulation, we generate a mixed phase space with regular islands surrounded by a chaotic sea (see Fig.~\ref{figure1}). The quantum mechanics of such a system leads to quantum states which are either localized on the regular islands or spread over the chaotic sea \cite{berryrobnik, Bohigas93}. This picture emerges in the semiclassical limit where the quantum scale represented by the Planck constant $\hbar$ is smaller than the typical area of the classical structures. Then, the regular islands form an effective lattice whose properties can be modified by varying the temporal modulation. More precisely, we demonstrate that, in such a driven lattice, the position and the number of regular islands as well as the size of the surrounding chaotic sea can be finely tuned. 

Another key phenomenon of purely quantum origin lies in the modification of the tunnel effect in such a mixed system. Indeed, quantum states can cross the classically forbidden dynamical barriers through a process similar to the standard tunneling called dynamical tunneling \cite{DynamicalTunneling, keshavamurthy2011dynamical}. This dynamical tunneling can be strongly affected by the presence of the chaotic states \cite{Tomsovic94, leyvraz1996level, bohigas1993quantum, Tomsovic_1998, PhysRevLett.84.4084, chepelianskii2002simulation, PhysRevE.82.056208, PhysRevE.64.025201, PhysRevLett.91.263601, peter0, brodier2002resonance, peter1, peter2, Ishikawa_2007, Shudo_2008, PhysRevLett.100.104101, Mertig_2013, PhysRevLett.104.114101, PhysRevE.96.040201}, which results in large variations of the tunneling rate over a short range of parameter, leading to so-called chaos-assisted tunneling resonances \cite{Tomsovic94, leyvraz1996level}. This effect has been experimentally observed with great precision for
classical waves \cite*{dembowski2000first, hofferbert2005experimental, PhysRevLett.100.174103, dietz2014spectral, PhysRevLett.104.163902, Kim:13}. For matter waves, dynamical and chaos-assisted tunneling between regular islands were achieved in the pioneering experiments \cite{Phillips,
  raizen, steck2002fluctuations}. However, experimental limitations had made it impossible to clearly observe the sharp resonances which are the hallmark of chaos-assisted tunneling \cite{Mouchet01, PhysRevLett.89.253201, PhysRevE.67.046216, PhysRevA.70.013408, DubertrandCAT}.

In this Letter, we show that in our ultra-cold atoms system we are able to experimentally produce an effective tunable superlattice and observe, for the first time, the
chaos-assisted tunneling resonances between spatially separated stable
islands, allowing us to control tunneling of cold atoms from full suppression to strong enhancement in a small range of parameters.

\textit{Experimental setup.--} Our experimental study is performed on a rubidium-87 Bose-Einstein condensate (BEC) machine, based on a hybrid trap \cite{FortunTunneltime}. We superimpose to the horizontal optical dipole beam ($X$-axis) of the trap a one-dimensional optical lattice obtained from the interference of two counter-propagating laser beams. We engineer the phase and amplitude of the lattice via three acousto-optic modulators \cite{CabreraSpectre}: one is dedicated to the control of the  lattice laser intensity and the others, driven by phase-locked synthesizers, control the relative phase, $\varphi$, between the two lattice beams. Introducing the dimensionless variables $p=2\pi P/(m \omega d)$, $x=2\pi X/d$ where $m$ is the atomic mass, $d=532$ nm is the lattice spacing, $X$ and $P$ are the position and momentum along the standing wave and normalizing the time $t$ to the modulation angular frequency $\omega$, the Hamiltonian that governs the dynamics reads \cite{Mouchet01, DubertrandCAT}
\begin{gather}
H=\frac{p^2}{2}-\gamma\left(1+\varepsilon \cos(t )\right)\cos \left( x + \varphi(t)\right) \; .
\label{eqH}
\end{gather}
Here, $\varepsilon$ is the amplitude of the lattice depth modulation and $\gamma=s\left(\omega_L/\omega\right)^2$, with the dimensionless depth $s$ of the lattice measured in units of the lattice characteristic energy $E_L=h^2/(2md^2)=\hbar \omega_L$. In practice, the dimensionless depth $s$ is precisely calibrated by monitoring the center-of-mass motion of the wave packets inside the lattice sites \cite{CabreraCalibration}. The quantum dynamics in that system is controlled by an effective tunable Planck constant $\heff=2\omega_L/\omega$, which gives the typical area occupied by a minimal wave packet in a dimensionless phase space \cite{Ott:ChaosDynSyst:02}. Here $\heff$ is thus an experimental parameter that effectively controls the quantum scale in the simulated system. It can be modified independently from $\gamma$ and $\varepsilon$, and thus for a fixed classical phase space.

\begin{figure}[h!]
\centering
\includegraphics{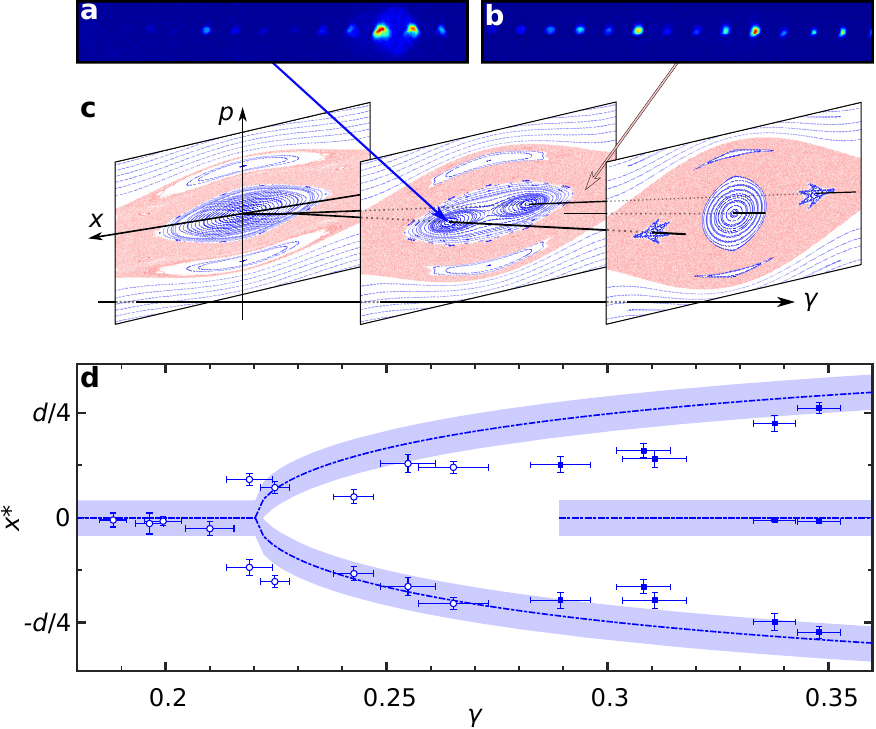}
\caption{\textbf{Observation of the bifurcation}. \textbf{a-b}, Typical images obtained after time-of-flight, when the wave packet lay initially in a regular island or in the chaotic sea respectively. In case (\textbf{a}), the momentum distribution remains narrow, while the wavefunction spreads over many orders in case (\textbf{b}). \textbf{c}, Evolution of the phase space for increasing values of the dimensionless depth $\gamma$ and fixed $\varepsilon=0.268$, showing the splitting of the initial central regular island into two, and then into three islands. \textbf{d}, Evolution of the position of the center of the regular islands, as a function of $\gamma$: experimental results obtained for $\hbar_{\textrm{eff}}=0.203$ (empty marks) and $\hbar_{\textrm{eff}}=0.232$ (filled marks). The analytical prediction is indicated as a dot-dashed line. The shaded area reveals the limited resolution governed by $\hbar_{\rm eff}$.}
\label{fig:bifurcation}
\end{figure}

To experimentally implement this Hamiltonian, we first load the BEC in a static lattice by a smooth increase of the lattice intensity \cite{CabreraSpectre}. We then apply the protocol \cite{DubertrandCAT} described in Fig.~\ref{figure1}:
the lattice is suddenly shifted at the desired position by an abrupt change of the phase $\varphi$ and its amplitude is subsequently modulated for an even number of modulation periods $n\times 2T$ (see below). In order to perform measurements on our system, we continue the modulation for an extra $T/2$ duration. This amounts to performing a $\pi/2$ rotation in phase space in order to transfer in momentum space the information encoded in position space (see SI) \cite{FortunTunneltime,DubertrandCAT}. After a 25 ms time-of-flight, the interferences between the matter waves originating from the different cells of the optical lattice create a diffraction pattern (see Figs.~\ref{figure1}e and \ref{fig:bifurcation}a$\&$b), from which the local population in the islands can be reconstructed (see Methods).

\textit{Classical dynamics and bifurcation.--}
The classical dynamics generated by the periodic Hamiltonian (\ref{eqH}) is made apparent by a stroboscopic phase space map, plotting the $(x,p)$ values at every modulation period. Typical $(x,p)$ spaces (phase spaces) are displayed in Fig.~\ref{figure1} (see also Fig.~\ref{fig:resonance}c) and show mixed dynamics with chaotic (red) and regular (blue) orbits. These orbits organize themselves in specific structures: regular islands made of stable periodic orbits, surrounded by a chaotic sea bounded in momentum by large-$p$ regular zones (whispering galleries). Qualitatively, the value of the modulation amplitude $\varepsilon$ mostly affects the size of the chaotic sea. By increasing $\gamma$ in the range $[0.18, 0.36]$ at fixed $\varepsilon= 0.268$, the initial central stable island splits into two, and even three, stable islands (see Fig.~\ref{fig:bifurcation}c) \cite{DubertrandCAT}.
This process is a Hopf type bifurcation well-known in dynamical systems \cite{leboeuf1999normal} (see SI). Beyond the first bifurcation, classical trajectories starting from one island reach the symmetric island at each period. As a result, classical atoms remain in the same island when observed every two periods. To highlight the quantum tunneling dynamics, we therefore probe the system after an \textit{even} numbers of modulation period.

In order to characterize such dynamical bifurcations in phase space, we apply the modulation for a short even number of periods (between 4 and 10) to a packet of atoms initially placed at different positions along the $x$-axis. Indeed, the dynamics at short time is essentially classical, as quantum interferences do not manifest themselves yet.
When the initial wave packet is placed into a stable island, we observe a diffraction pattern after time-of-flight with only a few orders populated (see Fig.~\ref{fig:bifurcation}a). This is to be contrasted with the pattern observed when atoms are placed into the chaotic sea, where many orders are populated (see Fig.~\ref{fig:bifurcation}b). 
The measurement of the standard deviation of the momentum distribution therefore provides information on the initial position (in a chaotic or a regular region) in phase space along the $x$ direction (see Methods).  More specifically,  the initial wave packet position $x_0$ that leads to a minimal momentum standard deviation corresponds to the center of the regular islands. The island positions measured experimentally are in global agreement with the analytical prediction \cite{DubertrandCAT} (see Fig.~\ref{fig:bifurcation}d). The major discrepancies appear close to the critical values $\gamma = 0.22$ and $\gamma = 0.29$, where bifurcations actually take place. Indeed, close to those bifurcation points, a relatively small distance separates the islands and we probe the classical bifurcation with a finite resolution dictated by the value of $\heff$ ranging from 0.2 to 0.23 for our experimental parameters. 

This splitting of the central island actually generates a synthetic stroboscopic superlattice with two characteristic lengths: the spacing $d=532$ nm of the original lattice, and the distance between the stable islands, tunable from 50 to 250 nm, in a given cell. The corresponding stroboscopic quantum dynamics is described in terms of eigenstates $\ket{\psi_n}$ of the Floquet operator $U_F$ (the evolution operator for two periods of modulation). Each Floquet eigenstate is associated with a quasi-energy $\varepsilon_n$ such that $U_F\ket{\psi_n}=\exp(-i\frac{\varepsilon_n 2T}{\heff})\ket{\psi_n}$. As a consequence the stroboscopic dynamics can be linked to an effective time-independent Hamiltonian with eigenstates $\ket{\psi_n}$ and associated energies $\varepsilon_n$.

\begin{figure}[h!]
\centering
\includegraphics{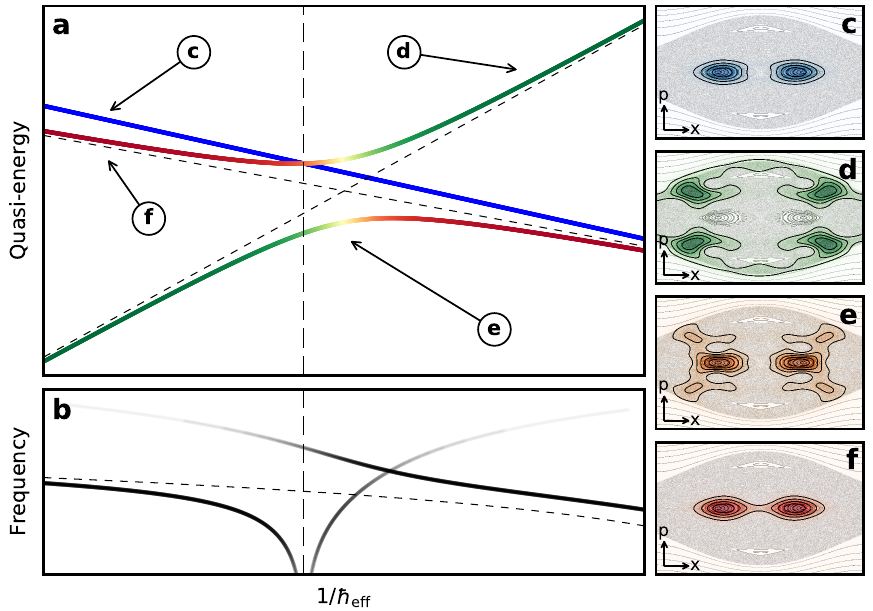}
\caption{\textbf{Sketch of a CAT resonance}. \textbf{a}, Quasi-energy spectrum of a regular doublet 
(red and blue) and a chaotic (green) state. An avoided crossing 
occurs between the chaotic (green) state and the regular 
state (red) part of the doublet having the same symmetry. As $\hbar_{\rm eff}$ is varied, these two states mix 
and repell. \textbf{b}, The resulting tunneling oscillation 
frequencies (proportional to differences in quasi-energies) exhibit two 
contributions and strong variations. \textbf{c-f}, Phase space representations of quantum states involved in the crossing (Husimi function, see \cite{husimi}).
Regular, antisymmetric (\textbf{c}) and symmetric (\textbf{f}), states are localized near integrable orbits; the chaotic states (\textbf{d}) lies in the chaotic sea while the mixed one (\textbf{e}) overlaps with both 
structures.}
\label{fig:principe-theorique}
\end{figure}

\textit{Chaos-assisted tunneling.--} To characterize chaos-assisted tunneling, we choose the regime for which two $x$-symmetric regular islands appear inside the chaotic sea. While a classical trajectory starting from a regular island will remain on the same island when observed every two periods (see above), a quantum wave packet initially
located on a stable island will oscillate between the two symmetric islands due to the tunneling effect.

The standard tunneling effect, as described in textbooks for regular systems, involves only a doublet of quasi-degenerate symmetric and antisymmetric states localized on two symmetric islands.
The tunneling period is proportional to the inverse of the splitting between these
two states.  In this case of regular dynamical tunneling (in the
absence of a chaotic sea), the splitting is a smooth function of the
parameters of the system, being an exponential of a classical action
divided by $\hbar$.

\begin{figure}[h!]
\centering
\includegraphics{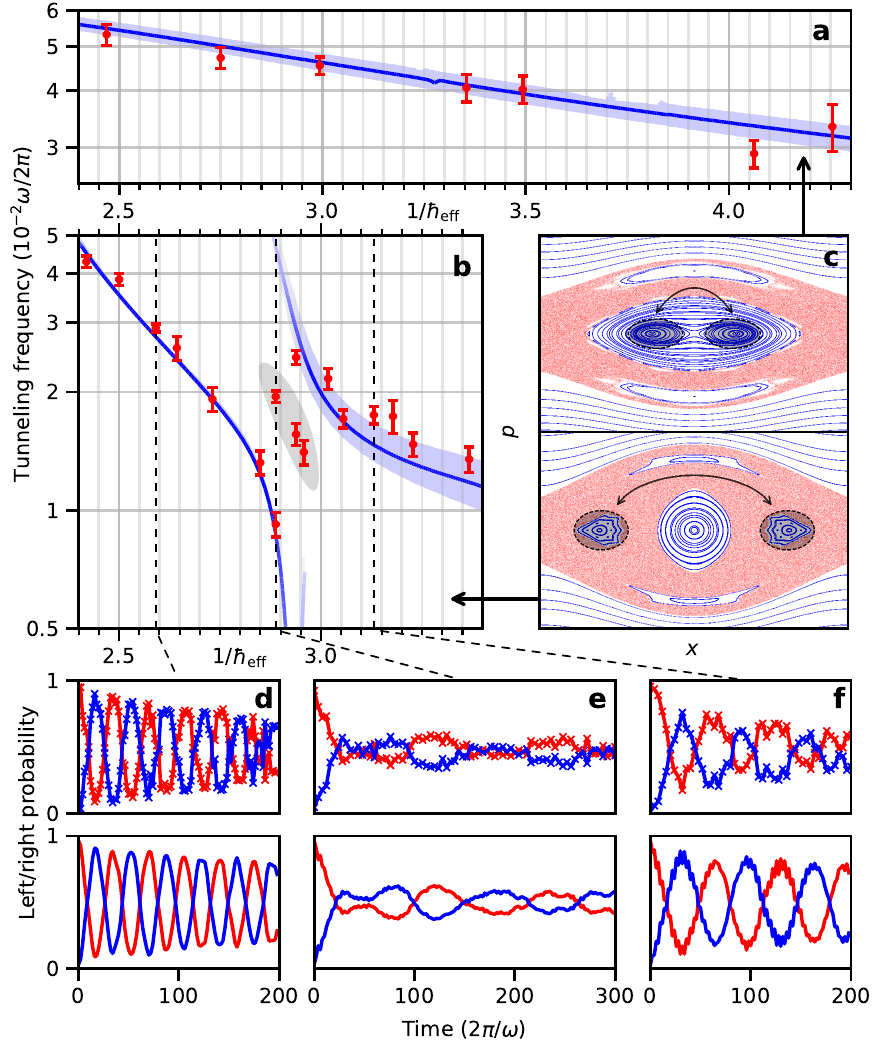}
\caption{\textbf{Experimental observation of a CAT resonance and comparison with regular dynamical tunneling}. 
\textbf{a-b} Experimentally measured tunneling frequencies (red dots) as a function of $\hbar_{\rm eff}^{-1}$ compared with numerical simulation in the case where all lattice cells are equally populated (blue lines) for the parameters $\varepsilon = 0.14$, $\gamma = 0.249\pm0.002$ (\textbf{a}) and $\varepsilon = 0.24$, $\gamma = 0.375  \pm 0.005$ (\textbf{b}). Tunneling frequencies are extracted through Fourier transform of the experimental or numerical oscillations (see Methods). Blue shaded area corresponds to the experimental uncertainty on $\gamma$, with a shade intensity giving the height of the Fourier peak. 
The grey shaded area of  (\textbf{b}) reveals the zone in which a good agreement with experiments requires realistic simulations that take into account the initial finite size of the BEC (see Methods). The corresponding classical phase spaces are plotted in \textbf{c}. \textbf{d-f}, Population in each island and for different values of  $\hbar_{\rm eff}^{-1}$  (corresponding to the dotted lines in panel (\textbf{b})) as a function of time (blue: left island, red: right island); upper panel: experimental data, lower panel: realistic simulations with 13 cells initially populated. The first three oscillations of \textbf{d} are represented in Fig.~\ref{figure1}\textbf{e}.}
\label{fig:resonance}
\end{figure}

The Figure~\ref{fig:principe-theorique} illustrates the mechanism of chaos-assisted tunneling.
In the presence of the chaotic sea, the tunnel
effect between the islands is generically mediated by an additional
chaotic state that interacts with one of the regular eigenstates
having the same symmetry (see Figs.~\ref{fig:principe-theorique}a and c-f) \cite{Tomsovic94}. This interaction causes an
avoided crossing between the regular and the chaotic levels, which in
turn can greatly modify the splitting between the two regular
levels (see Figs.\ref{fig:principe-theorique}a and b). As the number of chaotic states is large in the semiclassical
limit, these crossings are numerous and translate into large and
erratic resonances of the tunneling rate when varying a parameter.

In Fig.~\ref{fig:resonance}, the experimentally measured oscillation frequencies are plotted as a function of $1/\heff$, keeping all other parameters fixed, in two different phase space configurations. In the first configuration (regular dynamical tunneling), there is no chaotic sea between the two islands, and we experimentally observe a smooth exponential decay of the frequency as a function of $1/\heff$. The two islands are actually ``protected'' by the regular tori within which they lie, which dramatically reduces the possibility for the dynamics to be contaminated with a state from the chaotic sea. In the second configuration, the two islands are separated by a chaotic sea, and we observe large non-monotonic variations of the tunneling frequency. The experimental data agree very well with numerical simulations, with no fitting parameter. At the resonance, the tunneling oscillations exhibit a beating with two frequencies (see Fig.~\ref{fig:resonance}e). This phenomenon is a clear signature of the three-level mechanism of chaos-assisted tunneling. The apparent discrepancy observed at resonance with the simplest theory built from the translationally invariant case originates from the finite spatial extension of the atomic wave packet in the optical lattice. Realistic numerical simulations including such effects agree very well with the experimental data (see Fig.~\ref{fig:resonance} d-f). 

We have observed similar chaos-assisted resonances at different parameter values (see SI). The observation of a large number of oscillations was made possible thanks to a few experimental tricks: (1) the optical confinement of the BEC is made with far-off resonance lasers to avoid the deleterious effect of spontaneous emission, (2) gravity is compensated for by a horizontal guide superimposed to the lattice and (3) the BECs used in the experiments contain a few 10$^4$ atoms. Indeed, the very same experiment performed with a larger number of atoms ($> 10^5$) yields a strong damping of the tunneling oscillations (see SI). We attribute the damping of these oscillations to interaction effects.

\textit{Conclusion.--}
We have shown that the intermediate regime of temporal driving that we have studied allows to design tunable effective superlattices through phase space engineering. Moreover, we have demonstrated that the presence of chaos leads to chaos-assisted tunneling resonances between sites of the superlattice, that we have observed experimentally for the first time in a quantum system.

This new type of control through complexity via temporal driving is fairly generic. Interestingly enough, in the context of cold atoms, the tunneling resonances that we have demonstrated are species independent, in contrast to the well-known Feshbach resonances for interactions \cite{Feshbach}.

Remarkably, chaos-assisted tunneling should allow to probe models with a new kind of long-range hoppings for quantum simulation, not mediated by interactions  \cite{bloch1,bloch2,nagerl}. Indeed, chaos-assisted tunneling implies mediation by a state delocalized in the chaotic sea which surrounds all the lattice sites. As a result, it leads to long-range hoppings across the system. Such long-range models have been extensively studied theoretically in condensed matter, and exhibit a rich phenomenology such as multifractality (power-law random banded matrices \cite{EveMir08}), spin-glass physics \cite{PhysRevLett.35.1792}, high-$T_C$ superconductivity \cite{PhysRevB.78.060502}, etc. Our protocol offers for the first time a strategy to emulate such models with cold atoms. 

\onecolumngrid

\bibliographystyle{apsrev4-1}
\bibliography{refs_CAT.bib}

\vspace{0.5 cm}

\noindent {\bf Acknowledgements} \\  This work was supported by Programme Investissements
d'Avenir under the program ANR-11-IDEX-0002-02, reference
ANR-10-LABX-0037-NEXT, and research funding
Grant No. ANR-17-CE30-0024. M.A. acknowledges support
from the DGA (Direction G\'en\'erale de l'Armement), and N.D. support from R\'egion Occitanie and Universit\'e Paul Sabatier. We thank M. Burgher for experimental support. We thank P. Schlagheck and B. Peaudecerf for useful discussions.
Computational resources were provided by the facilities of Calcul en Midi-Pyr\'en\'ees (CALMIP).\\

\noindent {\bf Author contributions} \\
M.A., G.C., N.D., J.B. and D.G.-O conducted the experiments. M.M., B.G. and G.L. performed the theoretical analysis. All authors edited and reviewed the manuscript. O. G., D. U., B.G., G.L., J.B. and D.G.-O. supervised the work. \\

\noindent {\bf Competing interests} The authors declare no competing interests.

\end{document}